\documentstyle[12pt]{article}
\let\ssection=\section
\renewcommand{\section}{\setcounter{equation}{0}\ssection}

\newcommand\mathC{\mkern1mu\raise2.2pt\hbox{$\scriptscriptstyle|$}
        {\mkern-7mu\rm C}}
\newcommand{\mathR}{{\rm I\! R}}

\newcommand\Q{{\cal Q}}

\newcommand\K{{\cal K}}
\newcommand\id{{\rm id}}

\newcommand\ket[1]{\,|#1\rangle}
\newcommand\Ob[1]{{\rm Ob}(#1)}
\newcommand\Hom[1]{{\rm Hom}(#1)}

\newcommand\Ran[1]{{\rm Ran}\;#1}
\newcommand\Dom[1]{{\rm Dom}\;#1}
\newcommand\AF[1]{{\rm AF}(#1)}
\newcommand\AFi[1]{{\rm AF}^{{\rm in}}(#1)}
\newcommand\AFo[1]{{\rm AF}^{{\rm out}}(#1)}
\newcommand\eq[1]{Eq.\ (\ref{#1})}
\newcommand\eqs[2]{Eqs.\ (\ref{#1}--\ref{#2})}

\newcommand\mapright[1]{\smash{
        \mathop{\mbox{\large{$\;\longrightarrow\;$}}}\limits^{#1}}}
\newcommand\shortmapright[1]{\smash{
        \mathop{\mbox{\large{$\;\rightarrow\;$}}}\limits^{#1}}}

\begin{document}
\begin{titlepage}
\hspace{8truecm}Imperial/TP/2-03/15


\begin{center}
{\large\bf A New Approach to Quantising Space-Time:\\[6pt]
        III.\ State Vectors as Functions on Arrows}
\end{center}

\vspace{0.8 truecm}

\begin{center}
            C.J.~Isham\footnote{email: c.isham@imperial.ac.uk}\\[10pt]
            The Blackett Laboratory\\
            Imperial College of Science, Technology \& Medicine\\
            South Kensington\\
            London SW7 2BZ\\
\end{center}

\begin{abstract}
In \cite{IshQCT1_03}, a new approach was suggested for
quantising space-time, or space. This involved developing
a procedure for quantising a system whose configuration
space---or history-theory analogue---is the set of
objects, $\Ob\Q$, in a (small) category $\Q$. The quantum
states in this approach are cross-sections of a bundle
$A\leadsto\K[A]$ of Hilbert spaces over $\Ob\Q$. The
Hilbert spaces $\K[A]$, $A\in\Ob\Q$, depend strongly on
the object $A$, and have to be chosen so as to get an
irreducible, faithful, representation of the basic
`category quantisation monoid'. In the present paper, we
develop a different approach in which the state vectors
are complex-valued functions on the set of {\em arrows\/}
in $\Q$. This throws a new light on the Hilbert bundle
scheme: in particular, we recover the results of that
approach in the, physically important, example when $\Q$
is a small category of finite sets.

\end{abstract}
\end{titlepage}

\section{Introduction}\label{Sec:Introduction}
In \cite{IshQCT1_03} (hereafter referred to as {\bf I.})
a new approach was developed for quantising space-time or
space. This was based on the observation that such
entities can be regarded as the objects in a category
whose arrows are structure-preserving maps. This lead to
considering the general problem of finding the quantum
theory of a system whose configuration space is the set
of objects, $\Ob\Q$, in a small\footnote{A `small'
category is one in which the collections of objects and
arrows are genuine sets, rather than just classes.}
category $\Q$. A key idea in {\bf I.}\ is to use the
monoid $\AF\Q$ of arrow fields on $\Q$ as an analogue of
the collection of momentum variables in a normal quantum
theory. Here, an `arrow field' $X$ is defined to be an
association of an arrow $X(A)$ to each object $A\in\Ob\Q$
with the property that the domain of $X(A)$ is $A$.

The simplest scheme of this type is where the state
vectors are complex-valued functions $A\mapsto\psi(A)$ on
$\Ob\Q$. The key operators are then defined by
\begin{eqnarray}
        (\hat a(X)\psi)(A)&:=&\psi(\Ran X(A))
                    \label{aXsimple}\\[3pt]
        (\hat \beta\psi)(A)&:=&\beta(A)\psi(A)
                    \label{betasimple}
\end{eqnarray}
for all $A\in\Ob\Q$. Here, $X$ is any arrow field, and
$\beta$ is any element of the set $F(\Ob\Q,\mathR)$ of
all real-valued functions on $\Ob\Q$. We also define the
exponentiated operator $\hat V(\beta):=\exp-i\hat \beta$:
\begin{equation}
    (\hat V(\beta)\psi)(A):=e^{-i\beta(A)}\psi(A).
\end{equation}

The operators $\hat a(X)$, $X\in\AF\Q$, and $\hat
V(\beta)$, $\beta\in F(\Ob\Q,\mathR)$, satisfy the
equations
\begin{eqnarray}
    \hat a(X_2)\hat a(X_1)&=&\hat a(X_1\&X_2)\label{CQAaa}\\
    \hat V(\beta_1)\hat V(\beta_2)&=&\hat
    V(\beta_1+\beta_2)                  \label{CQAVV}\\
    \hat a(X)\hat V(\beta)&=&\hat V(\beta\circ\ell_X)\hat
    a(X)                                \label{CQAaV}
\end{eqnarray}
where `$\&$' denotes the multiplication operation (see
\eq{Def:X2X1}) in the monoid $\AF\Q$, and $\ell_X$ is the
action of the arrow field $X$ on the set of objects,
defined by $\ell_X A:=\Ran X(A)$ for all $A\in\Ob\Q$.

Equations (\ref{CQAaa}-\ref{CQAaV}) correspond to a
representation of the `category quantisation monoid' of
$\Q$, defined to be the semi-direct product
$\AF\Q\times_\ell F(\Ob\Q,\mathR)$. In general, the
possible quantisations of the system are deemed to be
given by the irreducible, faithful representations of
this monoid.

However, the simple representation given by
\eqs{aXsimple}{betasimple} is not faithful: in
particular, it fails to distinguish between arrows with
the same domain and range. More precisely, to any arrow
$f\in\Hom\Q$, there corresponds the arrow field $X_f$
\begin{equation}
    X_f(A):=\left\{ \begin{array}{ll}
            f &\mbox{\ if $\Dom f=A$;}
                                    \label{Def:Xf}\\[3pt]
                 \id_A & \mbox{\ otherwise,}
                 \end{array}
        \right.
\end{equation}
for all $A\in\Ob\Q$. A representation is then said to
`separate arrows' (or `distinguish arrows') if $\hat
a(X_f)\neq \hat a(X_g)$ for all arrows $f,g\in\Hom\Q$
with the same domain and range. This is manifestly false
in the case of \eq{aXsimple}. Thus we have an incomplete
representation of the categorial structure.

The main task in {\bf I.} was to develop a quantisation
scheme in which arrows {\em are\/} separated. This
involved choosing state vectors to be cross-sections of a
Hilbert bundle $\K$, $A\leadsto \K[A]$, on $\Ob\Q$, where
the Hilbert spaces $\K[A]$ (which are strongly
$A$-dependent) must be chosen so as to give an
irreducible, faithful, representation of the category
quantisation monoid. We will refer to this scheme as the
`Hilbert presheaf' approach since, as shown in {\bf I.},
$\K$ is actually a presheaf of Hilbert spaces.

For the special, and physically very important, case when
$\Q$ is a category of finite sets (for example, causal
sets), this scheme was discussed in detail in
\cite{IshQCT2_03} (hereafter referred to as {\bf II.})
where it was argued that an appropriate choice for the
Hilbert spaces $\K[A]$, $A\in\Ob\Q$, is\footnote{$|A|$
denotes the number of elements in the finite set $A$.}
$\mathC^{|A|}$.

In the present paper we return to this example and
rederive the results from a new perspective that further
enhances their plausibility. This is a special case of a
new approach in which the quantum states are
complex-valued functions on the set of {\em arrows\/} in
the category, rather than Hilbert-bundle valued functions
on the set of objects.

Having quantum states as functions on arrows has the
advantage that it is easier to explore the problem of
separating arrows with the same domain and range. In
fact, the scheme can be viewed as an extension of the
simple approach with complex-valued state functions
$A\mapsto\psi(A)$ since a function $A\mapsto \psi(A)$ can
be regarded as a special example of a function $h\mapsto
\phi(h)$ in which $\phi$ depends on only the {\em
ranges\/}\footnote{Alternatively, $\phi$ can depend only
on the {\em domain\/} of the arrows. However, choosing
$\phi$ to depend on the ranges of the arrows allows for
the neatest link with the ideas of {\bf I}.} of the
arrows $h\in\Hom\Q$.

The central idea is that by utilising aspects of an arrow
in addition to its range, we can obtain a satisfactory
quantum theory using only complex-valued state functions.
In particular, arrows with the same domain and
range---which are not separated by state vectors
$A\mapsto\psi(A)\in\mathC$---will be separated in the new
approach. Of course, we anticipate that there should be
some relation between this approach and the one based on
a presheaf of  Hilbert spaces. In the important example
of quantising on a category of sets we shall see
explicitly that this is indeed the case.

The plan of the paper is as follows. In Section
\ref{SubSec:ActionAFHomQ} we introduce the monoids of
`in' and `out' arrow fields and show how they act on the
set of objects in the category. This action is used in
Section \ref{SubSec:RepInOutAF} to construct a
representation of these monoids by operators acting on
the space of complex-valued functions on $\Hom\Q$. The
formalism is extended in Section \ref{SubSec:ConfigOps}
to include configuration space operators.

The ensuing representation of the category quantisation
monoid is highly reducible because of the action of the
`in' arrow fields and the action of a second copy of the
space of configuration functions. This issue is addressed
in Section \ref{Sec:ReducingArrows} by considering how to
reduce the number of arrows on which the state vectors
are defined. We start in Section
\ref{SubSec:RewritingFormalism} by re-expressing the
formalism in a way which emphasises that the state
vectors can be written in the form $\psi(A,h)$ where
$A\in\Ob\Q$, and $h$ is any arrow whose range is $A$.
This is used in Section \ref{SubSec:IntroPreArrows} to
motivate the introduction of a presheaf of arrows on $\Q$
which must be chosen  such that the representation of the
category quantisation monoid becomes irreducible, while
remaining faithful (that is, while still separating
arrows with the same domain and range).

As shown in Section \ref{SubSec:PresheafGlobalElements},
if the category $\Q$ has a terminal object $1$, a natural
choice for this presheaf is to associate with each object
$A$ the set of arrows $h:1\rightarrow A$; in other words,
the set of global elements of $A$. By this means, we are
able to reproduce the results obtained in {\bf II.}\
(using a presheaf of Hilbert spaces) for the case when
$\Q$ is a category of finite sets.

\section{Using The Monoid $\AFi\Q\times\AFo\Q$}
\label{Sec:UsingMonoidAFiF}
\subsection{The Action of $\AFi\Q\times\AFo\Q$ on $\Hom\Q$}
\label{SubSec:ActionAFHomQ} In the first two papers in
this series, {\bf I.} and {\bf II.}, only `out' arrow
fields were used, where an `out' arrow field $X$ is
defined to be an association to each object $A\in\Ob\Q$,
of an arrow $X(A)$ whose domain is $A$. However, there is
obviously another type of arrow field (an `in' field)
which associates to each object $A$ an arrow $Y(A)$ whose
{\em range\/} is $A$. We shall denote by $\AFi\Q$ and
$\AFo\Q$ the set of all `in', and `out', arrow fields
respectively.

In {\bf I.}, the set $\AFo\Q$ was given a monoid
structure\footnote{The unit element is the arrow field
$\iota$ defined by $\iota(A):=\id_A$ for all
$A\in\Ob\Q$.} by defining the product of
$X_1,X_2\in\AFo\Q$ as
\begin{equation}
     X_2\& X_1(A):=X_2(\Ran{X_1(A)})\circ X_1(A)
     \label{Def:X2X1}
\end{equation}
for all $A\in\Ob\Q$. In diagrammatic form, if
$A\shortmapright{X_1(A)}B\shortmapright{X_2(B)}C$, then
\begin{equation}
        X_2\& X_1(A):= X_2(B)\circ X_1(A).
\end{equation}

Similarly, if $Y_1,Y_2\in\AFi\Q$, we define
\begin{equation}
    Y_2\& Y_1(D):= Y_1(D)\circ Y_2(\Dom{Y_1(D))}
    \label{Def:Y2Y1}
\end{equation}
for all $D\in\Ob\Q$. In diagrammatic form, if
$F\shortmapright{Y_2(E)}E\shortmapright{Y_1(D)}D$, then
\begin{equation}
        Y_2\& Y_1(D):= Y_1(D)\circ Y_2(E).
\end{equation}

The Hilbert presheaf quantisation scheme discussed in
{\bf I.} and {\bf II.} is based on the action of the
monoid $\AFo\Q$ on the set $\Ob\Q$ given by
\begin{equation}
    \ell_X(A):=\Ran{X(A)}\label{Def:ellXA}
\end{equation}
for all $X\in\AFo\Q$, $A\in\Ob\Q$. Thus, if
$X(A):A\rightarrow B$, then $\ell_X(A)=B$. This is an
`action' in the sense that (i) for each $X\in\AFo\Q$,
$\ell_X$ is a transformation of $\Ob\Q$ to itself; and
(ii) for all $X_1,X_2\in\Ob\Q$, we have
$\ell_{X_2}\circ\ell_{X_1}=\ell_{X_2\&X_1}$.

The `in' arrow fields also act on $\Ob\Q$, with
\begin{equation}
        \varrho_Y(A):=\Dom{Y(A)}\label{Def:varrhoA}
\end{equation}
for all $Y\in\AFi\Q$, $A\in\Ob\Q$. This might suggest
that the product monoid $\AFi\Q\times\AFo\Q$ acts on
$\Ob\Q$. However, this is not so since in general
$\varrho_Y(\ell_X(A))\neq \ell_X(\varrho_Y(A))$, and
hence the obvious transformations induced by pairs
$(Y,X)\in\AFi\Q\times\AFo\Q$ on $\Ob\Q$ do not represent
the monoid structure of $\AFi\Q\times\AFo\Q$. This is why
`in' vector fields were not introduced into the quantum
scheme described in {\bf I.}\ and {\bf II.}

However, the situation changes if we let the state
functions be defined on arrows, rather than objects. For
there {\em is\/} an obvious action of
$\AFi\Q\times\AFo\Q$ on $\Hom\Q$, namely
\begin{equation}
        h\mapsto X(\Ran{h})\circ h\circ Y(\Dom{h})
        \label{YXactonh}
\end{equation}
for all $h\in\Hom\Q$ and $(Y,X)\in \AFi\Q\times\AFo\Q$.
In considering this expression it may be helpful to use
the diagram
\begin{equation}
    E\mapright{Y(D)}D\mapright{h}R\mapright{X(R)}B.
\end{equation}

\subsection{Representing the `In' and `Out' Arrow Fields}
\label{SubSec:RepInOutAF} With the quantum state vectors
now being taken as functions
$\phi:\Hom\Q\rightarrow\mathC$, we use the action in
\eq{YXactonh} to represent the arrow fields as follows.
If $X\in\AFo\Q$, then we define (cf.\ \eq{aXsimple})
\begin{eqnarray}
     (\hat a(X)\phi)(h)&:=&\phi(X(\Ran h)\circ h)
     \label{Def:a(X)phi(h)}\\
          &=&\phi(l_{X(\Ran{h})}h)
\end{eqnarray}
and if $Y\in\AFi\Q$, then
\begin{eqnarray}
        (\hat l(Y)\phi)(h)&:=&\phi(h\circ Y(\Dom{h}))\\
        &=&\phi(r_{Y(\Dom{h})}h)
\end{eqnarray}
where the actions of $\AFo\Q$ and $\AFi\Q$ on $\Hom\Q$
are defined from the transformation in \eq{YXactonh} as
\begin{eqnarray}
            l_X(h)&:=& X(\Ran{h})\circ h\\[2pt]
            r_Y(h)&:=&    h\circ Y(\Dom{h})
\end{eqnarray}
for all $h\in\Hom\Q$, $X\in\AFo\Q$ and $Y\in\AFi\Q$.

To see that these operators form genuine (anti-)
representations of the appropriate monoids, suppose first
that $X_1$ and $X_2$ are a pair of `in' arrow fields
which, for expository clarity, satisfy the diagram
\begin{equation}
    D\mapright{h}R\mapright{X_1(R)}B
    \mapright{X_2(B)}C.
\end{equation}
Then
\begin{eqnarray}
        (\hat a(X_1)\hat a(X_2)\phi)(h)
        &=&(\hat a(X_2)\phi)(X_1(R)\circ h)\nonumber\\
        &=&\phi\left(X_2(B)\circ(X_1(R)\circ h)\right)
\end{eqnarray}
for all $h\in\Hom\Q$. On the other hand,
\begin{eqnarray}
    (\hat a(X_2\&X_1)\phi)(h)=\phi\left((X_2(B)\circ
    X_1(R))\circ h\right)
\end{eqnarray}
and thus, using the associativity of arrow composition in
the category $\Q$, we see that
\begin{equation}
        \hat a(X_1)\hat a(X_2)=\hat a(X_2\&X_1)
        \label{aX1aX2=}
\end{equation}
as claimed.

Similarly, if $Y_1,Y_2\in\AFi\Q$, then, with the aid of
the diagram
\begin{equation}
    F\mapright{Y_2(E)}E\mapright{Y_1(D)}D\mapright{h}R
\end{equation}
we see that
\begin{eqnarray}
        (\hat l(Y_1)\hat l(Y_2)\phi)(h)
            &=&(\hat l(Y_2)\phi)(h\circ Y_1(D))\nonumber\\
            &=&\phi\left((h\circ Y_1(D))\circ Y_2(E)\right).
\end{eqnarray}
On the other hand, \eq{Def:Y2Y1} gives
$(Y_2\&Y_1)(D)=Y_1(D)\circ Y_2(E)$, and so, for all
$h\in\Hom\Q$,
\begin{equation}
        (\hat l(Y_2\&Y_1)\phi)(h)=
                \phi\left(h\circ(Y_1(D)\circ Y_2(E))\right).
\end{equation}
Hence, for all $Y_1,Y_2\in\AFi\Q$,
\begin{equation}
    \hat l(Y_1)\hat l(Y_2)=\hat l(Y_2\& Y_1)
           \label{lY1lY2=}
\end{equation}
as claimed.

Finally, with the aid of the diagram
\begin{equation}
E\mapright{Y(D)}D\mapright{h}R\mapright{X(R)}B
\end{equation}
we see that, for all $X\in\AFo\Q$, $Y\in\AFi\Q$, and
$h\in\Hom\Q$,
\begin{eqnarray}
(\hat a(X)\hat l(Y)\phi)(h)&=&
        (\hat l(Y)\phi)(X(R)\circ h)\nonumber\\
        &=&\phi\left((X(R)\circ h)\circ Y(D)\right)\nonumber\\
        &=&\phi\left(X(R)\circ (h\circ Y(D))\right)\nonumber\\
        &=&\left(\hat a(X)\phi\right)(h\circ Y(D))\nonumber\\
        &=&(\hat l(Y)\hat a(X)\phi)(h)
\end{eqnarray}
and hence
\begin{equation}
        \hat a(X)\hat l(Y)=\hat l(Y)\hat a(X)
        \label{aXlY=}
\end{equation}
for all $X\in\AFo\Q$ and $Y\in\AFi\Q$.

From \eq{aX1aX2=}, \eq{lY1lY2=} and \eq{aXlY=} it follows
that we have an operator (anti-) representation of the
monoid $\AFi\Q\times\AFo\Q$.

\subsection{The Configuration Space Operators}
\label{SubSec:ConfigOps} The next step is to represent
the configuration-space functions
$\beta:\Ob\Q\rightarrow\mathR$. In the quantisation
scheme based on sections of a bundle of Hilbert spaces
$A\leadsto \K[A]$, a function $\beta\in F(\Ob\Q,\mathR)$
is represented by the operator $(\hat\beta\psi)(A)
:=\beta(A)\psi(A)$ or, in exponentiated form,
\begin{equation}
    (\hat V(\beta)\psi)(A):=e^{-i\beta(A)}\psi(A)
\end{equation}
and we need an analogue of this.

In the present context, there are two natural ways of
representing a function $\beta$. Namely
\begin{equation}
        (\hat\rho(\beta)\phi)(h):=\beta(\Ran{h})\phi(h)
        \label{Def:rhobeta}
\end{equation}
and
\begin{equation}
        (\hat\lambda(\beta)\phi)(h):=\beta(\Dom{h})\phi(h).
        \label{Def:lambdabeta}
\end{equation}

Then, for all $h\in\Hom\Q$, $X\in\AFo\Q$ (and denoting
$\Ran h$ by $R$ for typographical clarity), we have,
using \eq{Def:a(X)phi(h)} and \eq{Def:rhobeta},
\begin{eqnarray}
    (\hat a(X)\hat\rho(\beta)\phi)(h)
        &=&(\hat\rho(\beta)\phi)(X(R)\circ h)\nonumber\\
    &=&\beta(\Ran{\{X(R)\circ h\}})\phi(X(R)\circ h)
    \nonumber\\
    &=&\beta(\Ran{X(R)})\phi(X(R)\circ h)
            \label{rXrhobeta=}
\end{eqnarray}
while, on the other hand,
\begin{eqnarray}
    (\hat\rho(\beta)\hat a(X)\phi)(h)
        &=&\beta(\Ran h)(\hat a(X)\phi)(h)\nonumber\\
        &=&\beta(R)\phi(X(R)\circ h).
        \label{rhobetarX=}
\end{eqnarray}
Now, if the function $\beta$ is replaced with the
function $\beta\circ\ell_X:\Ob\Q\rightarrow\mathR$, the
right side of \eq{rhobetarX=} now contains $\beta(\ell_X
R)$. But,  by the definition of $\ell_X$, we have
$\ell_XR=\Ran{X(R)}$, and hence
\begin{equation}
    \hat a(X)\hat\rho(\beta)=
            \hat\rho(\beta\circ\ell_X)\hat a(X).
\end{equation}
Thus,  if $\hat V(\beta):=\exp-i\hat\rho(\beta)$, we have
a representation of the category quantisation monoid
$\AFo\Q\times_\ell F(\Ob\Q,\mathR)$.

Next, we must consider the implications of the second
beta operator, $\hat\lambda(\beta)$, defined in
\eq{Def:lambdabeta}. If $Y\in\AFi\Q$ then, for all
$h\in\Hom\Q$ (and denoting $\Dom{h}$ by $D$ for
typographical clarity), we have
\begin{eqnarray}
    (\hat\lambda(\beta)\hat l(Y)\phi)(h)
    &=&\beta(D)(\hat l(Y)\phi)(h)\nonumber\\
    &=&\beta(D)\phi(h\circ Y(D))\label{lambdabetalphi}
\end{eqnarray}
while
\begin{eqnarray}
    (\hat l(Y)\hat\lambda(\beta)\phi)(h)
    &=&(\hat\lambda(\beta)\phi)(h\circ Y(D))\nonumber\\
    &=&\beta(\Dom{[h\circ Y(D)]})\phi(h\circ
                                    Y(D))\nonumber\\
    &=&\beta(\Dom{Y(D)})\phi(h\circ Y(D)).
    \label{lambdabetaphi}
\end{eqnarray}
However, the action of the `in' arrow field $Y$ on
$\Ob\Q$ is $\varrho_Y:A\mapsto\Dom{Y(A)}$, and hence,
from \eq{lambdabetalphi}, it follows that, for all
$h\in\Hom\Q$
\begin{eqnarray}
    (\hat\lambda(\beta\circ\varrho_Y)\hat l(Y)\phi)(h)
    &=&(\beta\circ\varrho_Y)(D)\phi(h\circ Y(D))
                        \nonumber\\
    &=&\beta(\Dom{Y(D)})\phi(h\circ Y(D)).
\end{eqnarray}
Thus
\begin{equation}
        \hat\lambda(\beta\circ\varrho_Y)\hat l(Y)=
        \hat l(Y)\hat\lambda(\beta)
\end{equation}
which, defining $\hat
W(\beta):=\exp-i\hat\lambda(\beta)$, corresponds to an
operator representation of the monoid
$\AFi\Q\times_\varrho F(\Ob\Q,\mathR)$.

Furthermore, if $X\in\AFo\Q$ and $\beta\in
F(\Ob\Q,\mathR)$, then (again with $h:D\rightarrow R$)
\begin{eqnarray}
    (\hat a(X)\hat\lambda(\beta)\phi)(h)
    &=&(\hat\lambda(\beta)\phi)(X(R)\circ h)\nonumber\\
    &=& \beta(\Dom{X(R)\circ h})\phi(X(R)\circ h)
                        \nonumber\\
    &=& \beta(D)\phi(X(R)\circ h).
\end{eqnarray}
However,
\begin{eqnarray}
    (\hat\lambda(\beta)\hat a(X)\phi)(h)
    &=&\beta(D)(\hat a(X)\phi)(h)\nonumber\\
    &=&\beta(D)\phi(X(R)\circ h)
\end{eqnarray}
and so
\begin{equation}
        [\,\hat a(X),\hat\lambda(\beta)\,]=0.
        \label{Comm:r(X)lambda(beta)}
\end{equation}
Of course, this could have been expected since the action
of $\hat a(X)$ involves attaching arrows to the range of
$h$, whereas $\hat\lambda(\beta)$ acts only on the domain
of $h$.

Similarly, it is easy to show that
\begin{equation}
[\,\hat \rho(Y),\hat l(\beta)\,]=0.
\label{Comm:l(Y)rho(beta)}
\end{equation}
In addition, it is obvious that, for all
$\beta_1,\beta_2\in F(\Ob\Q,\mathR)$,
\begin{equation}
    [\,\hat\rho(\beta_1),\hat\lambda(\beta_2)\,]=0
\end{equation}
which, together with \eq{aXlY=},
\eq{Comm:r(X)lambda(beta)} and \eq{Comm:l(Y)rho(beta)},
implies that we have an operator representation of the
monoid $(\AFo\Q\times_\ell F(\Ob\Q,\mathR))\times
    (\AFi\Q\times_\varrho F(\Ob\Q,\mathR))$.

\section{Reducing the Number of Arrows}
\label{Sec:ReducingArrows}
\subsection{Rewriting the Formalism}
\label{SubSec:RewritingFormalism} The representation
constructed above of the category quantisation monoid
$\AFo\Q\times_\ell F(\Ob\Q,\mathR)$ certainly separates
the arrows of $\Q$. Indeed, if $f,g:A\rightarrow B$, then
(cf.\ \eq{Def:Xf})
\begin{equation}
    (\hat a(X_f)\phi)(\id_A)=\phi(f)
\end{equation}
and
\begin{equation}
    (\hat a(X_g)\phi)(\id_A)=\phi(g)
\end{equation}
which, since there are certainly functions $\phi$ such
that $\phi(f)\neq \phi(g)$, shows that $\hat a(X_f)\neq
\hat a(X_g)$.

However, there are two problems with this representation.
The first is that in order to construct a proper Hilbert
space for the quantum states
$\phi:\Hom\Q\rightarrow\mathC$, we need a measure $\mu$
on $\Hom\Q$, so that the inner product between two states
$\phi$ and $\psi$ can be defined as
\begin{equation}
\langle\phi,\psi\rangle:=\int_{\Hom\Q}d\mu(h)\,\phi(h)^*\psi(h).
\end{equation}
This is as challenging as the corresponding problem in
the Hilbert presheaf scheme of finding measures on
$\Ob\Q$. If $\Hom\Q$ is finite, or countably infinite,
the obvious discrete measures can be used. In other
situations, a more detailed investigation is needed, and
I hope to return to this in a later paper.

The second problem is that because the quantum states
carry a representation of the monoid $(\AFo\Q\times_\ell
F(\Ob\Q,\mathR))\times
    (\AFi\Q\times_\varrho F(\Ob\Q,\mathR))$
even if this representation is irreducible, the
associated representation of the category quantisation
monoid $\AFo\Q\times_\ell F(\Ob\Q,\mathR)$ will generally
not be.

One obvious solution is to reduce the number of arrows on
which the state functions are defined, and to help in
this task it is useful to think of the states we have
introduced as functions $\phi(R,h)$ where $R\in\Ob\Q$ and
$h\in\Hom{{\rm -},R}$, the set of all arrows whose range
is $R$. In this notation, the operators defined above are
as follows, where $X\in\AFo\Q$, $Y\in\AFi\Q$,
$\beta_1,\beta_2\in F(\Ob\Q,\mathR)$:
\begin{eqnarray}
(\hat a(X)\phi)(R,h)&:=&\phi(B,X(R)\circ h)
    \label{Def:rXphiAh}\\
(\hat\rho(\beta_1)\phi)(h)&:=&\beta_1(R)\phi(R,h)
    \label{Def:rhobetaphiAh}\\
(\hat l(Y)\phi)(R,h)&:=&\phi(R,h\circ Y(D))
    \label{Def:lyphiAh}\\
(\hat\lambda(\beta_2)\phi)(R,h)&:=&\beta_2(D)\phi(R,h)
    \label{Def:lambdabetaphiAh}
\end{eqnarray}
for all $R\in\Ob\Q$ and $h\in\Hom{{\rm -},R}$. Once
again, for typographical clarity we are using the diagram
\begin{equation}
    E\mapright{Y(D)}D\mapright{h}R\mapright{X(R)}B.
        \label{Diagram:EDAB}
\end{equation}

Of course, the `$R$' label is not really necessary, since
the arrow $h$ already `knows' what its range is. However,
by writing the operator definitions in this form it is
clear that the action of the `out' arrow fields is on the
objects in $\Q$---as in the Hilbert presheaf scheme---but
now supplemented by an action on the arrows that come
into each object. It is also clear how the extra
operators ({\em i.e.}, those outside the category
quantisation monoid $\AFo\Q\times_\ell F(\Ob\Q,\mathR)$)
$\hat l(Y)$ and $\hat \lambda(\beta)$ act on the `left
hand end' of the arrow $h\in\Hom{{\rm -},R}$. It is the
existence of these operators that is responsible for the
reducible nature of the representation of
$\AFo\Q\times_\ell F(\Ob\Q,\mathR)$.

\subsection{Introducing Presheaves of Arrows}
\label{SubSec:IntroPreArrows} As remarked above, to get
an irreducible representation of $\AFo\Q\times_\ell
F(\Ob\Q,\mathR)$ the number of arrows that enter each
object must be restricted, albeit in such a way that the
quantisation scheme still separates arrows. Thus we are
interested in subsets of arrows $J[R]\subset \Hom{{\rm
-},R}$ so that we can define $\hat a(X)$ as (cf.\
\eq{Def:rXphiAh})
\begin{equation}
(\hat a(X)\phi)(R,h):=\phi(\Ran{X(R)},X(R)\circ h)
    \label{Def:rXphiAhJ}
\end{equation}
for all $h\in J[R]$, $R\in\Ob\Q$. Clearly, for
\eq{Def:rXphiAhJ} to be consistent we require that if
$h\in J[R]$ and $f:R\rightarrow B$, then $f\circ h\in
J[B]$.

To understand better what this condition means, note that
the association $A\leadsto \Hom{{\rm -},A}$ defines a
covariant presheaf  {\bf H} on $\Q$; {\em i.e.}, {\bf H}
is a covariant functor from the category $\Q$ to the
category of sets {\bf Set}. More precisely:
\begin{enumerate}
\item[i)] to each object $A$ in $\Q$, the functor {\bf H}
associates the set $\Hom{{\rm -},A}$ of all arrows whose
range is $A$;
\item[ii)] to each arrow $f:A\rightarrow B$, the functor {\bf H}
associates the function $H(f):\Hom{{\rm -},A}\rightarrow \Hom{{\rm
-},B}$ defined by
\begin{equation}
        H(f)(h):=f\circ h
\end{equation}
for all $h\in \Hom{{\rm -},A}$.
\end{enumerate}

Then the condition above on the sets $J[A]\subset
\Hom{{\rm -},A}$, $A\in\Ob\Q$, means that the association
$A\leadsto J[A]$ corresponds to a presheaf ${\bf J}$ that
is a {\em subobject\/} of ${\bf H}$ in the category
$\Q^{\rm Set}$ of (covariant) presheaves on $\Q$.

In the (topos) category $\Q^{\rm Set}$, the sub-objects
of a presheaf {\bf H} are in one-to-one correspondence
with arrows in the category $\Q^{\rm Set}$
\begin{equation}
    \chi_J: \makebox{\bf H}\rightarrow
    \makebox{\boldmath$\Omega$}
    \label{chiJ}
\end{equation}
where {\boldmath$\Omega$} is the presheaf of {\em
sieves\/} on $\Q$. Here, we recall that a sieve on an
object $A\in\Ob\Q$ is defined to be a set $S$ of arrows
in $\Q$ whose domain is $A$ (thus $S\subset \Hom{A, {\rm
-}}$) and with the property that if $f\in S$ then $k\circ
f\in S$ for all arrows $k$ such that $\Dom{k}=\Ran{f}$
(thus, in a poset category, a sieve is just an upper set)
\cite{MM92}. We denote by $\Omega[A]$ the set of all
sieves on $A$. If $f:A\rightarrow B$, the map
$\Omega(f):\Omega[A]\rightarrow \Omega[B]$ is defined by
\begin{equation}
    \Omega(f)(S):=\{g\in \Hom{B, {\rm -}}\mid
    g\circ f\in S\}
\end{equation}
for all $S\in \Omega[A]$.

The arrow $\chi_J$ in \eq{chiJ} associates to each
$A\in\Ob\Q$, a map $\chi_J[A]:\Hom{{\rm
-},A}\rightarrow\Omega[A]$ which is defined by saying
that if $h\in \Hom{{\rm -},A}$ then $\chi_J[A](h)$ is the
set of all arrows $k$ with domain $A$ such that $k\circ
h$ belongs to $J[\Ran{k}]$. It is easy to see that these
form a sieve on $A$.

The challenge is to find a presheaf {\bf J} of arrows
such that the associated representation of the category
quantisation monoid is irreducible whilst maintaining the
separation of arrows in the quantum scheme. Such a
presheaf can be used in two ways:
\begin{enumerate}
    \item  {\bf J} can be used as it stands.
    The states are then functions $\phi(R,h)$, where $h\in
    J[R]$ is an arrow whose range is $R\in\Ob\Q$.

    \item However, if desired certain arrows can be added to the
    domain of the state vectors in addition to those in
    the subsets $J[A]$, $A\in\Ob\Q$. To see this, for
    each object $R$ consider the arrows $h\in\Hom{{\rm -},R}$ that
    are `picked out' by
    $\chi_J[R]$ in the sense that there exists
    $B\in\Ob\Q$ and an arrow $k:R\rightarrow B$, such that
    $k\circ h\in J[B]$.

    Now, using Dirac notation, when $\hat a(X)^\dagger$ acts on a state $\ket{B,m}$,
    $m\in \Hom{{\rm -},B}$ it takes it into a linear combination
    of states $\ket{R,h}$, $h\in \Hom{{\rm -},R}$, with the property that (i)
    $X(R):R\rightarrow B$; and (ii) $X(R)\circ h =m$.\footnote{This
    is the analogue of the results on the adjoints of operators
    obtained in {\bf I.}}  If
    the extra arrows `picked out' by $\chi_J$ are added to the domain of the
    state vectors, then if $m\in J[B]$ this sum would include
    {\em all\/} the states $\ket{R,h}$ ($R\in\Ob\Q$
    $h\in \Hom{{\rm -},R}$) such that $\chi_J[R](h)=m$.

\end{enumerate}

\subsection{The Presheaf of Global Elements of the Objects of
$\Q$} \label{SubSec:PresheafGlobalElements} We will now
discuss one particularly important example of a presheaf
{\bf J}. This is motivated by remarking, once again, that
the representation of the category quantisation monoid
$\AFo\Q\times_\ell F(\Ob\Q,\mathR)$ is reducible because
of the existence of the operators defined in
\eq{Def:lyphiAh} and \eq{Def:lambdabetaphiAh} as
\begin{eqnarray}
(\hat l(Y)\phi)(R,h)&:=&\phi(R, h\circ Y(D))
        \label{lYphiAh2}\\[3pt]
(\hat\lambda(\beta)\phi(R,h)&:=&\beta(D)\phi(R,h)
        \label{lambdabetaphiAh2}
\end{eqnarray}
where once again we have used the diagram
\begin{equation}
E\mapright{Y(D)}D\mapright{h}R\mapright{X(R)}B
\end{equation}
for typographical clarity.

One way of trivialising these operators that latch onto
the `left hand end' of the arrows $h$ is available if
there is a {\em terminal object\/}\footnote{A terminal
object in a category $\Q$ is an object $1$ with the
property that, for each object $A$ in $\Q$, there is
precisely one arrow $1_A:A\rightarrow 1$. It is easy to
see that any two terminal objects are isomorphic.} $1$ in
the category $\Q$. In this case, for all $R\in\Ob\Q$, we
can try restricting the domains of the arrows in $J[R]$
to be $1$, so that each arrow $h\in J[R]$ is of the form
$h:1\rightarrow R$. In other words, $h$ is a {\em
global\/} element of the object $R$. Thus $J[R]$ is
defined to be the set of all global elements of $R$. Note
that, if $h:1\rightarrow R$ is a global element of $R$,
then $X(R)\circ h:1\rightarrow B$, and hence $X(R)\circ
h$ is a global element of $B$. Thus the association
$A\leadsto J[A]$ defines a genuine presheaf {\bf J} that
is a subobject of the presheaf {\bf H}.

This choice for {\bf J} works because, since there is
only one arrow from $1$ to itself ({\em i.e.}, ${\rm
id}_1$) the only `in' arrow field that can act on an
arrow $h:1\rightarrow R$, is the identity. Similarly, the
operator $\hat\lambda(\beta)$  is now
\begin{equation}
        (\hat\lambda(\beta)\phi)(R,h)=\beta(1)\phi(R,h)
\end{equation}
for all $R\in\Ob\Q$ and $h\in J[R]$. This can safely be
ignored since it just multiplies each state vector by the
same complex number.

Thus we are left with the operators
\begin{eqnarray}
(\hat a(X)\phi)(R,h)&:=&\phi(B,X(R)\circ h)\\[3pt]
(\hat V(\beta)\phi)(R,h)&:=&e^{-i\beta(R)}\phi(R,h)
\end{eqnarray}
for all $R\in\Ob\Q$, and all $h:1\rightarrow R$. This is
certainly a representation of the category quantisation
monoid, and so the important question now is whether it
separates arrows. Clearly this will be so provided, if
$f,g:R\rightarrow B$, there exists $h:1\rightarrow R$
such that $f\circ h\neq g\circ h$.

The existence of such separating arrows depends on the
category. For example, if $\Q$ is a category of sets then
everything works. This is because the global elements of
a set are just its elements in the normal sense---so that
$J[A]\simeq A$ for all $A\in\Ob\Q$---and, of course, a
function between two sets is completely specified by its
action on these elements. Thus, in this situation, the
quantum states are functions $\phi(A,a)$, where $a\in A$,
and basic operators are
\begin{eqnarray}
(\hat a(X)\phi)(A,a)&:=&\phi(B,X(A)(a))
            \label{Def:aXphia}\\[3pt]
(\hat V(\beta)\phi)(A,a)&:=&e^{-i\beta(A)}\phi(A,a)
\label{Def:Vbetaphia}
\end{eqnarray}
for all $a\in A$, and where $X(A)$ is now a function from
the set $A$ to the set $B$.

However,  \eqs{Def:aXphia}{Def:Vbetaphia} essentially
reproduce the Hilbert presheaf representation discussed
in detail in {\bf II}. There, the Hilbert space $\K[A]$
was identified with $\mathC^{|A|}$, $A\in\Ob\Q$, and this
is isomorphic to the space of complex-valued functions on
$A$. Furthermore, the representation of $X\in\AFo\Q$ by
the operator $\hat a(X)$ in \eq{Def:aXphia} reproduces
the representation by the corresponding operator in the
presheaf of Hilbert spaces. Thus the approach developed
above further justifies the choice  made in {\bf II.} of
the Hilbert presheaf for quantising on a category of
sets.

In fact, these considerations suggests a general way of relating
the structure in the present paper with that developed in {\bf I.}
and {\bf II}. Specifically, if {\bf J} is a presheaf of arrows,
the state vectors are functions $\phi(R,h)$ where $R\in\Ob\Q$ and
$h\in J[R]\subset\Hom{{\rm -},R}$. Thus each fixed $R$ gives a
function $\phi_R:J[R]\rightarrow\mathC$ defined as
$\phi_R(h):=\phi(R,h)$ for all $h\in J[R]$. Now, for each
$A\in\Ob\Q$, let $\K[A]$ be the set, $F(J[A],\mathC)$, of all
complex-valued functions on $J[A]$; and if $f:A\rightarrow B$
define $\kappa(f):\K[B]\rightarrow\K[A]$ by
\begin{equation}
    (\kappa(f)v)(h):=v(f\circ h)
\end{equation}
for all $v\in F(J[B],\mathC)$, and $h\in J[A]$. It is
easy to check that these assignments of $\K[A]$,
$A\in\Ob\Q$, and $\kappa(f):\K[B]\rightarrow\K[A]$
constitute a presheaf of vector spaces over $\Q$. Thus we
regain the structure discussed in {\bf I.}\ and {\bf II.}
Of course, we need to worry about putting inner products
on these vector spaces, but that is closely related to
the question of measures on $\Hom\Q$, which is a subject
for future research.

\section{Conclusion}

We have introduced an approach to quantising on a
category that provides an alternative to the techniques
used in the earlier papers {\bf I.} and {\bf II}. The new
scheme involves defining quantum states to be
complex-valued functions on the set of arrows in the
category $\Q$, rather than cross-sections of a
bundle/presheaf of Hilbert spaces over the set of objects
in $\Q$.

Arguably, in some respects this new approach is more
transparent  than the original one; in any event, it
certainly provides a new way of looking at things. For
example, it throws light on the choice of Hilbert-space
presheaves used in the original approach; particularly
when $\Q$ is a category of sets. In general, in the new
approach we need to find presheaves of arrows, which in
some respects is a simpler thing to do. For a category of
structured sets (for example: topological spaces; causal
sets), this may help with the problem of finding
presheaves that distinguish only the structure-preserving
maps between sets.

From a mathematical perspective, one of the main problems
is to find suitable measure structures on the space of
arrows $\Hom\Q$ in $\Q$; or on the appropriate subspace
of $\Hom\Q$ if a sub-sheaf of arrows is selected. This is
not dissimilar to the problem of constructing measures on
the set of objects, $\Ob\Q$, and is something to which I
hope to return later.

\section*{Acknowledgements}

\noindent Support by the EPSRC in form of grant GR/R36572
is gratefully acknowledged. I am very grateful to Jeremy
Butterfield for a critical reading of the manuscript.

\end{document}